\documentclass[sigconf]{acmart}

\usepackage{booktabs}
\usepackage{caption}	 

\usepackage{paralist}

\usepackage{balance}	
\usepackage{graphicx}
\makeatletter
\def\maxwidth{\ifdim\Gin@nat@width>\linewidth\linewidth
\else\Gin@nat@width\fi}
\makeatother

\AtBeginEnvironment{quote}{\small}
\AtEndEnvironment{quote}{}

	\author{Wouter Groeneveld}
			\orcid{https://orcid.org/0000-0001-5099-7177}
		\affiliation{
					\institution{KU Leuven}
									\city{Leuven}
							\country{Belgium}
			}
	\email{wouter.groeneveld@kuleuven.be}
	\author{Lynn Van den Broeck}
		\affiliation{
					\institution{KU Leuven}
									\city{Leuven}
							\country{Belgium}
			}
	\email{lynn.vandenbroeck@kuleuven.be}
	\author{Joost Vennekens}
		\affiliation{
					\institution{KU Leuven}
									\city{Leuven}
							\country{Belgium}
			}
	\email{joost.vennekens@kuleuven.be}
	\author{Kris Aerts}
		\affiliation{
					\institution{KU Leuven}
									\city{Leuven}
							\country{Belgium}
			}
	\email{kris.aerts@kuleuven.be}

\renewcommand{\shortauthors}{Pre-print of accepted ITiCSE 2022 paper.}

\keywords{creativity; self-assessment; creative problem solving} 

\ccsdesc[500]{Social and professional topics~Student assessment}
\ccsdesc[500]{Social and professional topics~Software engineering education}

\begin{document}

	\title{Self-Assessing Creative Problem Solving for Aspiring Software
Developers: A Pilot Study}

\begin{abstract}
 We developed a self-assessment tool for computing students in higher education to measure their Creative Problem Solving skills. Our survey encompasses 7 dimensions of creativity, based
on existing validated scales and conducted focus groups. These are: \emph{technical knowledge}, \emph{communication},
\emph{constraints}, \emph{critical thinking}, \emph{curiosity},
\emph{creative state of mind}, and \emph{creative techniques}. Principal
axis factor analysis groups the dimensions into three overarching
constructs: \emph{ability}, \emph{mindset}, and \emph{interaction}. The
results of a pilot study (\(n\) = 269) provide 
evidence for its psychometric qualities, making it a useful instrument for educational researchers to investigate students' creative skills.
\end{abstract}

\maketitle
\bibliographystyle{ACM-Reference-Format}

\hypertarget{introduction}{%
\section{Introduction}\label{introduction}}

\label{sec:intro}

Industry experts perceive \emph{creativity} as an important non-technical skill for software engineers  \citep{groeneveld2020non}. Nevertheless, current curricula pay relatively little explicit attention to this skill \citep{groeneveld2020soft}.  This may in part be due to the fact that creativity is a complex and broad concept \citep{davis1999barriers}, which has not yet been widely researched in the field of Software Engineering (SE) \citep{amin2017snapshot}.

In this paper, we are interested in the concept of \emph{creative problem solving} (CPS) , i.e., the use of creativity
to solve day-to-day programming problems.  We build on previous work, that examined this concept by means of a focus group study \citep{groeneveld2021exploring}. This work, which we will further refer to simply as \emph{the FGS},  identified 39 themes that are relevant to CPS, grouped in seven domains: \emph{technical knowledge},
\emph{communication}, \emph{constraints}, \emph{critical thinking},
\emph{curiosity}, \emph{creative state of mind}, and \emph{creative techniques}. Due to page limits, we cannot give more details about the FGS here, but they are available in \citep{groeneveld2021exploring}. Our goal is to make the insights from the FGS actionable for computing education research. In particular, we derive from this study a creativity self-assessment  tool for SE students, called the \emph{Creative Programming Problem Solving Test} (CPPST). It allows students to measure their skills in each of the seven creativity domains.  This may be useful for students themselves, e.g., as a means of tracking
their progress throughout a curriculum, but it is mainly intended as an instrument for educational researchers who wish to investigate the effect that certain changes to the curriculum or to a particular course have on the students' creative skills.

The remainder of this paper is divided as follows.
Section \ref{sec:bg} outlines related work, while
Section \ref{sec:method-design} and \ref{sec:method-analysis} describe
the used methodology to design and analyze the survey. Next,
in Section \ref{sec:results} and \ref{sec:disc}, we present and discuss
the results, followed by possible limitations in Section
\ref{sec:limit}. Finally, we conclude this work in Section
\ref{sec:concl}.

\hypertarget{background}{%
\section{Background}\label{background}}

\label{sec:bg}

In general, creativity is often measured through divergent thinking,  as is the case for the popular Torrance's Test of
Creative Thinking (TTCT) \citep{torrance1972predictive}. While the TTCT can be used  to assess \emph{creative potential} \citep{kim2020longitudinal}, it does not attempt to measure the \emph{creative process} itself \citep{miller2014self}.
To address this, \citet{miller2014self} developed the Cognitive
Processes Associated with Creativity (CPAC) scale, focusing on the processes of brainstorming, metaphorical and analogical
thinking, perspective-taking, imagery, incubation, and flow.  The FGS also identified these topics, but, in addition, the industry professionals also stressed topics such as constraint-based thinking
and communication (see Section \ref{sec:method-design}). In the SE context, the CPAC scale is therefore only part of the story.

The same can be said for similar efforts of measuring creativity in general, such as the Creativity Support Index (CSI) of 
\citet{carroll2009creativity}. It  consists of these components: exploration, expressiveness, immersion,
effort/results, enjoyment, collaboration. This list also lacks several important topics from the FGS. However, the \emph{expressiveness} component was not mentioned in the FGS: while self-expression is often thought to be a key part of creativity in general, industry experts do not appear to consider it as an important component of CPS in software engineering.

Scales have also been developed to measure creativity specifically in engineering
students.  \citet{amato2011assessing} based their scale on three existing
scales: the Resistance to Change scale, the Curiosity and Exploration
Inventory, and the Zampetakis \& Moustakis Scale that assesses
entrepreneurial intentions. The FGS did not mention  entrepreneurial
intentions, and while general engineering and
SE share many conceptual ideas, they differ radically in execution.
There is no ability to apply a bugfix patch to a badly engineered
bridge, for example. Their exploratory factor analysis revealed four
main factors: cognitive approaches (the usage of problem solving
tactics, mapping to our CPS domain \emph{creative techniques}),
cognitive challenges (the enjoyment of experiences that challenge
viewpoints; this somewhat maps to \emph{curiosity}), cognitive
preparedness, and impulsivity in problem solving. The last two
components are personality-based, which our CPS model does not focus on.

There also exist multiple scales that focus on computational thinking.
The Computational Thinking
Scale (CTS) of \citet{korkmaz2017validity} contains the following five factors: creativity,
cooperativity, algoritmic thinking, critical thinking and problem
solving. Although the CTS scale confirms the importance of a few domains
that our CPS framework shares, such as the presence of creative and
communicative skills, it lacks important aspects from the FGS, such as  the influence of constraints and the usage of practical creative
techniques. To measure their single ``creativity'' factor,  they selected suitable items from the ``How Creative Are You?'' scale,
the ``Problem Solving Scale'', the ``Cooperative Learning Attitude
Scale'', ``The Scale of California Critical Thinking Tendency'', and the
``Logical-Mathematical Thinking'' scale.

In summary, although self-assessment metrics exist for creativity in general, in an engineering context  and in the context of computational thinking, none fully suits our needs.  Indeed, existing scales focus on other aspects of creativity than CPS and do not cover all of the aspects identified by industry experts in the FGS. 

\hypertarget{survey-design}{%
\section{Survey Design}\label{survey-design}}

\label{sec:method-design}

For each of the seven domains identified in the FGS, we composed eight questions, based on existing surveys. An overview of the full questionnaire is
available in Appendix \ref{sec:appendix}. Our intention is to reduce the
item set after determining the internal consistency by inspecting
Cronbach's \(\alpha\) values. Bad questions will be omitted in future versions of the survey. The analysis is described in Section \ref{sec:method-analysis}.

Since we are interested in CPS in the context of SE students, we designed the scale to be as domain-specific as
possible, as recommended by \citet{barbot2019creativity}. Therefore, we replace generic questions such as ``I like
variety'' by, e.g.: ``I liked varied aspects of the programming
assignment''.

Assessment tools for creativity can be divided into those measuring creative
potential, performance, and achievement \citep{barbot2019measuring}. Our CPPST 
measures \emph{achievements}. It does not gauge creative potential, nor
does it measure performance, as Torrance's TTCT or Amabile's Consensual
Assessment Technique (CAT) \citep{amabile1988model} metrics do. The
items are self-assessing in nature and ask about the usage of various
CPS skills on a recent programming achievement.

All questions were sampled on a Likert-5 scale, which is  the most
common in creativity research
\citep{kaufman2012counting}. Some surveys on which we based our
questions use a Likert-6 scale with ``not applicable'' as the sixth
option, but we left this out as all items are deemed
relevant. The answer categories are: (1) completely disagree, (2)
disagree, (3) neutral, (4) agree, (5) completely agree. Two questions per domain are reverse-coded to check the consistency of the answers. For each survey, questions are served in random order. Students were not able to see the results after finishing the questionnaire. 

We now describe the origins of the questions for each of the seven FGS domains.

\hypertarget{technical-knowledge-knw}{%
\subsection{Technical Knowledge (KNW)}\label{technical-knowledge-knw}}

To be able to creatively solve programming problems, one needs to have technical knowledge and to invest in keeping this knowledge up-to-date, by continuously learning and improving oneself. In the FGS, the following themes emerged w.r.t.~technical
knowledge: \emph{continuous learning}, \emph{domain models},
\emph{seeking out different inputs}.

To assess these themes, we adapt the following 8 questions from the 14-item lifelong learning
measurement scale of \citet{kirby2010development}:

\begin{compactenum}
\def\labelenumi{\arabic{enumi}.}

\item
  I have gained little knowledge during the project.
  \emph{\small(inverted)\normalsize}
\item
  I learned and applied new practical programming techniques.
\item
  I have gained insight into the problem domain.
\item
  The technical aspect of programming appealed to me.
\item
  I thought about my learning process and how to improve it.
\item
  I felt uncomfortable with this project because many aspects were
  unknown. \emph{\small(inverted)\normalsize}
\item
  I tried to relate the new knowledge to something I know.
\item
  Thanks to the project I also gained knowledge of other things outside
  of programming.
\end{compactenum}

\hypertarget{communication-com}{%
\subsection{Communication (COM)}\label{communication-com}}

The FGS stresses that modern SE is a team-based activity. Engaging in discussions helps tremendously in figuring out how to solve difficult programming problems. According to the FGS, communication is therefore an important
aspect of creative work in SE.

To assess this, we combined results from the FGS with ideas from the Collaboration Self
Assessment Tool \citep{ofstedal2009collaboration}. The latter survey is based on the
following concepts: (1) \emph{Contributions}, (2) \emph{Quality of
Work}, (3) \emph{Time management}, (4) \emph{Team Support}, (5)
\emph{Preparedness}, (6) \emph{Problem solving}, (7) \emph{Team
Dynamics}, (8) \emph{Interactions with Others}, (9) \emph{Role
Flexibility}, and (10) \emph{Reflection}. Since some of these concepts
are already present in the other domains of our CPPST scale, we 
focus here on communication, collaboration, and team efforts.

Related themes from the FGS are: \emph{rubber ducking} (to explain a problem to oneself or others to gain new insights), \emph{drawing on a whiteboard},
\emph{getting fast feedback}, \emph{working with peers},
\emph{responsibility}. The following questions assess communication in
the CPPST:

\begin{compactenum}
\def\labelenumi{\arabic{enumi}.}

\item
  I hardly asked for feedback from my fellow students.
  \emph{\small(inverted)\normalsize}
\item
  I visualized the problem on a whiteboard or on paper.
\item
  I regularly asked feedback from my teachers.
\item
  I supported my teammates by helping them with their tasks.
\item
  My own tasks were not completed on time so teammates ran into problems
  with the deadline. \emph{\small(inverted)\normalsize}
\item
  I supported the ideas and efforts of my teammates.
\item
  I was so proud of our result that I showed it to everyone.
\item
  I thoroughly thought suggestions by others through.
\end{compactenum}

\hypertarget{constraints-ctr}{%
\subsection{Constraints (CTR)}\label{constraints-ctr}}

Constraints outline the context of the problem. They keep you from
working on it forever, and they keep the solution relevant to the issue
at hand. Most software projects are ``brownfield'' projects, in which an
existing code base comes with many external constraints. Thus, both designing
and thinking in context of constraints are very important in software
engineering.

\citet{biskjaer2020task} recently demonstrated that there is a sweet spot
for accelerating (the potential for) creativity with constraints. Too few
constraints cause one to lose interest, while too many constraints may cause stress and hamper creative
freedom. According to \citet{biskjaer2013self}, self-imposed creativity
constraints can also play an important role. For our survey, however, we
view such self-imposed constraints as a creative technique, and
focus here on external constraints.

Themes that emerged from the FGS are: \emph{client-oriented designing}, \emph{efficiency},
\emph{relevance}, \emph{context}, \emph{performance},
\emph{time-constraints}. The following questions assess the ability to
effectively handle constraints:

\begin{compactenum}
\def\labelenumi{\arabic{enumi}.}

\item
  I regularly thought about the correctness of my solution.
\item
  Due to time pressure, I performed less well.
  \emph{\small(inverted)\normalsize}
\item
  I tried to make my code as elegant as possible.
\item
  I tried to identify the constraints of the assignment.
\item
  I have had the program tested by friends and / or family.
\item
  There was too much creative freedom for me, so I could not make a good
  decision. \emph{\small(inverted)\normalsize}
\item
  Coding on short notice accelerated my learning process.
\item
  I regularly tested myself and paid attention to its ease
  of use.
\end{compactenum}

\hypertarget{critical-thinking-cri}{%
\subsection{Critical Thinking (CRI)}\label{critical-thinking-cri}}

Critical thinking is the ability to question things, to come up
with alternatives, and to judge the trustworthiness of information
sources. A classic counterexample in the SE world is
mindlessly copy-pasting code snippets from online sources.
 \citet{sosu2013development} developed and validated a Critical Thinking Disposition Scale. Their 
scale is split into two main components after an exploratory factor
analysis: (A) Critical Openness and (B) Reflective Scepticism. We
adapted an equal number of items from both components to our domain.

Themes that emerged from the FGS
are: \emph{coming up with alternatives}, \emph{asking `why?'}. The
following questions assess the ability to think critically in the CPPST:

\begin{compactenum}
\def\labelenumi{\arabic{enumi}.}

\item
  In discussions about problems, I often suggested alternatives.
\item
  I regularly carefully weighed up the various options we had.
\item
  I dared to completely rewrite my code when it didn't go well.
\item
  I used multiple sources to find out information myself.
\item
  I didn't think it was important to ask teammates how they implemented
  something. \emph{\small(inverted)\normalsize}
\item
  I always check the credibility of the source when I look something up.
\item
  It was more important that it worked than that I 100\% understood why.
  \emph{\small(inverted)\normalsize}
\item
  Looking at other projects made me think about my own.
\end{compactenum}

\hypertarget{curiosity-cur}{%
\subsection{Curiosity (CUR)}\label{curiosity-cur}}

As the FGS states, a ``hungry'' mind, that is motivated to seek out new things, is very important in SE. Themes that emerged in the FGS are:
\emph{getting out of the comfortzone}, \emph{motivation}, \emph{the
creative urge}, \emph{complexity is fun}, \emph{admiration}. We base our questions about this domain on the Curiosity Index developed by
\citet{fulcher2004towards}:

\begin{compactenum}
\def\labelenumi{\arabic{enumi}.}

\item
  During the project, I got very much out of my comfort zone.
\item
  Many parts of the project piqued my curiosity.
\item
  I enjoyed getting involved in many aspects of the project.
\item
  I enjoyed really immersing myself in some aspects.
\item
  I was stimulated by the complexity of the project.
\item
  I felt the urge to implement extras.
\item
  I have not had any fun while developing the project.
  \emph{\small(inverted)\normalsize}
\item
  I had to commit myself to finish the project.
  \emph{\small(inverted)\normalsize}
\end{compactenum}

\hypertarget{creative-state-of-mind-mnd}{%
\subsection{Creative State of Mind
(MND)}\label{creative-state-of-mind-mnd}}

Optimal creative work requires a certain state of mind,
which  \citet{csikszentmihalyi1997flow} calls \emph{flow}.
The FGS mentions that people can tell whether colleagues are being creative by looking at body language: Are they happy, and making a lot of
jokes? Are they `in the zone'?

\citet{jackson1996development} developed  the Flow State Scale to measure Optimal
Experience, based on the work of Csikszentmihalyi. It contains 36 items
divided into nine factors: (1) Challenge-Skill Balance, (2)
Action-Awareness Merging, (3) Clear Goals, (4) Unambiguous Feedback, (5)
Concentration on Task, (6) Sense of Control, (7) Loss of
Self-Consciousness, (8) Transformation of Time, and (9) Autotelic
Experience.

Themes that emerged from the FGS are: \emph{focus}, \emph{flow/being in the zone},
\emph{flexibility}, \emph{environment (space/social)},
\emph{productivity tooling}. The common divisor of these themes and the
factors of the Flow State Scale leads to the following
questions:

\begin{compactenum}
\def\labelenumi{\arabic{enumi}.}

\item
  I remained focused for a long time on one part of the project.
\item
  I used productivity tools to focus more on the essentials
  (e.g.~shortcuts, command line tools, \ldots).
\item
  I found the experience to be very rewarding.
\item
  Time seemed to fly while working.
\item
  I did not know enough to meet the high demands of the project.
  \emph{\small(inverted)\normalsize}
\item
  Programming went almost automatically.
\item
  I did not know what exactly I wanted to achieve.
  \emph{\small(inverted)\normalsize}
\item
  I was not concerned with what others thought of my code.
\end{compactenum}

\hypertarget{creative-techniques-tch}{%
\subsection{Creative Techniques (TCH)}\label{creative-techniques-tch}}

Creative techniques are concrete actions to
accelerate CPS.  \citet{hunt2008pragmatic} describes various creative techniques specifically for software
developers. The FGS also mentioned other simple tricks, such as taking a coffee break or
deliberately going to the toilet.

We were unable to find an existing scale that measures the use of such techniques. Therefore, we based the questions solely on the following themes from the FGS:
 \emph{viewing problems from a different angle (birds-eye
view)}, \emph{asking energizer questions}, \emph{brainstorming},
\emph{mapping problems from different domains}, \emph{combining ideas},
and \emph{seeking out input}. The following questions assess these:

\begin{compactenum}
\def\labelenumi{\arabic{enumi}.}

\item
  I always used the same method to solve a problem.
  \emph{\small(inverted)\normalsize}
\item
  I used knowledge from another domain to solve something.
\item
  I combined different ideas to tackle a problem.
\item
  I deliberately took occasional breaks to let things sink in.
\item
  I brainstormed with others to come up with new ideas.
\item
  I took a step back now and then to see things as a whole.
\item
  In case of problems I let myself be inspired by other projects.
\item
  I was regularly stuck \emph{\small(inverted)\normalsize}
\end{compactenum}

\hypertarget{face-validity}{%
\subsection{Face validity}\label{face-validity}}

\label{sec:design-crossv}

We believe that the face validity of our CPPST is by design already quite high, since it is based on existing validated scales and the FGS of \citep{groeneveld2021exploring}, in which 33 experts collaborated on the exploration of creativity in SE. Furthermore, the selected items all resemble real-life CPS behavior.

Throughout the initial development of the scale, we received feedback from the department of psychology at a neighboring university. One of their recommendations was to use eight items per domain, instead of five as we had originally planned. Once all of their feedback had been addressed, we administered the survey to a small set of volunteering second-year SE students (\(n\) = 9), which led us to reformulate some sentences that were hard to understand.
The survey was also checked by a number of the FGS participants (\(n\) = 5), which resulted in
rewriting six items and changing the Likert label for value 3 from
``more or less'' to ``neutral''.  The approval of the items by these members of the FGS 
 provides further evidence for the face validity of
the survey.

\hypertarget{survey-analysis}{%
\section{Survey Analysis}\label{survey-analysis}}

\label{sec:method-analysis}

We wanted the data to represent a broad range of SE students. Therefore, we surveyed both first-year (\(n\) = 141) and last-year (\(n\) = 128) engineering students. The test was administered as close as possible to a project deadline as part of a programming course. The survey was not obligatory, but students were strongly encouraged to participate. Because student demographics in the surveyed group were not very heterogeneous and \citet{miller2014self} earlier found no significant difference in self-assessment creativity scores based on participant gender, ethnicity or year in school, we decided not to consider demographics as a separate discriminant.

We follow reliability and validity analysis procedures from Miller
\citep{miller2014self} and Korkmaz et al. \citep{korkmaz2017validity}.
However, we will be conservative when deciding whether to drop an item from the
scale when factor analysis  shows it to be loaded on
multiple factors. Since the CPS domains are heavily intertwined, as also confirmed by the FGS, we believe that items loaded on multiple factors cannot be avoided entirely, without losing key information.  We will therefore be hesitant to do so, especially if the Kaiser-Meyer-Olkin statistic that indicates the factorability of the item set is high enough
\citep{kaiser1974index}.

For the reliability analysis, we calculated several Cronbach's
\(\alpha\) values: an initial $\alpha$ for the entire survey,
a separate $\alpha$ after a number of items were dropped, and $\alpha$ values for each discovered factor. The items that were dropped were those with doubtful or worse item-total correlations (\(R_{it}\)), in accordance with the
rules of thumb of Ebel and Frisbie \citep{ebel1972essentials}: poor
(\(R_{it} < .20\)), doubtful (\(.21 < R_{it} <.29\)), good
(\(.30 < R_{it} < .39\)), and very good (\(R_{it} >.40\)).

Construct validity is usually investigated
with the help of various factor analysis methods
\citep{ebel1972essentials}. 
Since our survey is based on the seven domains of the FGS, we first perform a confirmatory factor analysis (CFA) for these 
domains. This will answer the question of whether the original seven domains can be considered independent factors for our survey. If the answer is no, we will perform an  exploratory factor analysis (EFA) to discover more appropriate factors.

\hypertarget{results}{%
\section{Results}\label{results}}

\label{sec:results}

In total, 269 out of 336 (\texttt{80\%}) students
participated. On average, the survey took 9.5 minutes. Overall, an initial reliability analysis resulted in a good global $\alpha$ of \texttt{.89}. However, when we look at each domain separately, several domains contained a number of items that have a low item-total correlation, according to the rules of  \citet{ebel1972essentials}. After removing the doubtful $R_{it}$ correlations, reducing the number of items from the initial 56 to 37, the global $\alpha$ increased to \texttt{.91}.  Full reliability item statistics can be found in Table \ref{tab:itemstats} at the end of this paper.  

The Kaiser-Meyer-Olkin statistic
for the 37 remaining items is \texttt{.90}, a value that Kaiser calls ``marvelous'' for the factorability~\citep{kaiser1974index}. This is also a promising indication for sound construct validity.

We then executed a confirmatory factor analysis with the seven FGS
domains as factors.
The Comparative Fit Index (CFI) is \texttt{.76}. The Root Mean Square
Error of Approximation (RMSEA) is \texttt{.066}, for which the P-value
of \(RMSEA \leq 0.05\) was zero. According to Kline, an acceptable
conformity should report values between the ranges
\(.06 \leq RMSEA \leq .08\) and \(.90 \leq CFI \leq .96\)
\citep{kline2015principles}. The bad CFI indicates that the seven-factor model is not a great fit for the data and that, as may have been expected, the FGS domains are not sufficiently 
independent factors. Therefore, we shifted from confirmatory to
exploratory analysis.

\begin{figure}[h!]
  \centering
    \includegraphics[width=0.49\textwidth]{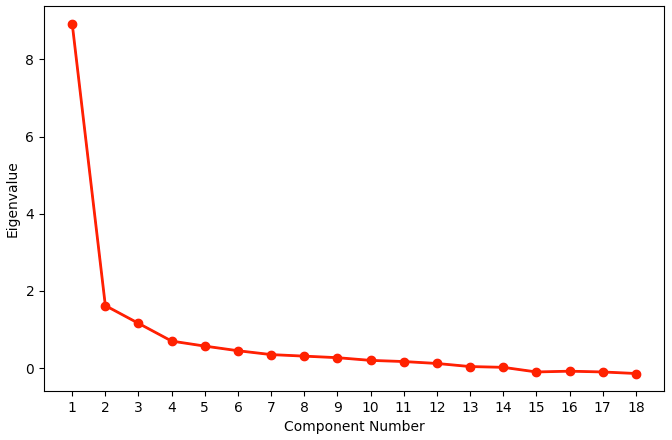}
    \caption{Scree plot (eigenvalues according to the factors). \label{fig:scree}}
\end{figure}

A parallel analysis suggests to use between three to five factors, as visible in the scree plot in Figure~\ref{fig:scree}. Five- and four-factor EFAs result in unsatisfactory component matrix distributions, while a three-factor EFA  with an oblique rotation reports a CFI of \texttt{.91} and
a RMSEA of \texttt{.06}.  We use the oblique rotation because we
recognize that there is likely to be some correlation between students'
latent subject matter preference factors. Table \ref{tab:fa} shows the three-factor loadings with principal axis factoring.

As could be expected from the lower $R_{it}$ correlations in Table \ref{tab:itemstats}
for domains such as \emph{critical thinking} (CRI) and 
\emph{creative techniques} (TCH), items related to those domains result
in low loading values in Table \ref{tab:fa}. For instance, only three CRI
items remain in the component matrix when employing a cutoff rate of
\texttt{.30}. A reliability analysis on the three revealed factors results in very good Cronbach's $\alpha$ values: \texttt{.89}, \texttt{.86} and \texttt{.83} for F1, F2, and F3 respectively. We further discuss these results in Section \ref{sec:disc}.

\begin{table}[h!]
  \centering
  \small
  \caption{Rotated Component Matrix: reduced 32-item CPPST set. Loadings below .30 (CRI7, TCH2, TCH5, CTR1, CTR2) are omitted for brevity. \label{tab:fa}}
  
    \newlength\oldwidth
    \setlength\oldwidth\arrayrulewidth
    \setlength\arrayrulewidth{1.2pt}

 \begin{tabular}{llccrllcc}
    \cline{1-4}     \cline{6-9} 
  \rule{0pt}{1.1em} Item & F1 & F2 & F3 & \rule{1cm}{0pt} &    Item & F1 & F2 & F3 \\
  \noalign{\global\setlength\arrayrulewidth{\oldwidth}}
    \cline{1-4}    \cline{6-9} 
    CUR2 &   .75 &      & & &          CUR7 &  .35 &    .47 &       \\
    CUR3 &   .51 &      & & &          CUR8 &   &       .32 &       \\
    CUR4 &   .70 &      & & &          MND5 &   &       .67 &       \\
    CUR5 &   .59 &      & & &          MND6 &   &       .67 &       \\
    CUR6 &   .56 &      & & &          TCH8 &   &       .76 &       \\
    MND3 &   .71 &      & & &          KNW6 &   &       .57 &       \\
        \cline{6-9} 
    MND4 &   .64 &      & & &          MND7 &   &            &  .42 \\
    TCH3 &   .44 &      & & &          COM4 &   &            &  .50 \\
    TCH6 &   .34 &      & & &          COM5 &   &            &  .42 \\
    KNW1 &   .68 &      & & &          COM8 &   &            &  .51 \\
    KNW2 &   .52 &      & & &          CTR3 &   &            &  .33 \\
    KNW3 &   .58 &      & & &          CTR8 &   &            &  .42 \\
    KNW4 &   .46 &  .41 & & &          CRI1 &   &            &  .47 \\
    KNW7 &   .35 &      & & &          CRI2 &   &            &  .31 \\
    KNW8 &   .40 &      & & &          CRI3 &   &            &  .36 \\
    COM7 &   .50 &      & \\       
    CTR7 &   .49 &      & \\       
\noalign{\global\setlength\arrayrulewidth{1.2pt}}

    \cline{1-4}     \cline{6-9} 
   \end{tabular}

\end{table}

\hypertarget{discussion}{%
\section{Discussion}\label{discussion}}

\label{sec:disc}
In this section, we will interpret the results of the pilot study in context of Creative Problem Solving and the FGS, as well as reflect on possible uses and future work. 

\hypertarget{revealed-constructs}{%
\subsection{Revealed constructs}\label{revealed-constructs}}

The factor analysis in Table \ref{tab:fa} shows that, except KNW4 and CUR7, the items are very evenly distributed across the three factors, which are deemed reliable thanks to the high $\alpha$ values. Factor F1 contains most Curiosity items, almost all
Knowledge, and Creative Techniques items, and some Creative State
of Mind items. These 17 items make up for the majority of the question
pool. It is clear that this group is knowledge oriented. Furthermore,
knowledge and curiosity-related questions show particularly high
loadings. We call this construct \emph{Ability}. To be able to solve a
programming problem creatively, one first needs to be able to program. Next to programming skills, knowledge of creative techniques that speed up the process are a big help. Curiosity may be a key motivation for investing the time and effort in acquiring all of these skills. KNW4, ``the technical aspect of programming appealed to me'', is loaded on both F1 and F2. Perhaps  some students are mesmerized by the programming challenge while others remain indifferent---independent of their technical ability.

Second, factor F2 contains Creative State of Mind items and a number of related items. For instance, TCH8, the highest loaded item, states ``I
was regularly stuck'', which is also related to one's mindset on how to
approach a problem. The feeling of being comfortable with
unknown aspects (KNW6) is also linked to one's belief in own abilities. CUR7, ``I had fun while developing the project'', is also loaded on both F1 and F2. This could perhaps indicate that, as Figueiredo et al. mentioned, while having fun is stimulating to learning to program, ``a lack of success on the (introductory) programming courses, can also be a demotivating factor''~\citep{figueiredo2019predicting}. A minimum level of ability could increase the coding fun. 

As F2 items are related to a way of thinking, we call F2 \emph{Mindset}. It seems related to
what Duckworth calls ``Grit'' or what Dweck calls a ``Growth Mindset''
\citep{hochanadel2015fixed}. In a recent study on the
relation between mindset in computing students and their study
performance, \citet{apiola2020mindset} found that  mindset on computing was growth-oriented, but that mindset on creativity was the most fixed of all scales. This means that students either think they are creative, or that they are not, but they are not open to the idea of
nurturing creativity. A tool such as our CPPST may be used to track how students' creativity increases over the course of a curriculum, thereby helping to cure students of this misconception. 

Lastly, factor F3 contains almost all communication-related items.
Other items, such as ``in discussions about problems, I often suggested
alternatives'' (CRI1) and ``I paid attention to its ease of use'' (CTR8)
also involve a interactive element. Therefore, we call F3
\emph{Interaction}. Social aspects have already been shown to be a critical component of CPS \citep{amato2011assessing}.  External suggestions,
support, and criticism are important to build creative
solutions and to build them quickly. Creative programmers
know how and how often to interact, and how to critically interpret
responses.  \citet{katz1984collective} talks about ``collective problem solving'' and discusses its many advantages over individual problem solving, which seem to translate well into the field of SE.

\hypertarget{firstyearvslastyear}{%
\subsection{First-year versus last-year students}\label{firstyearvslastyear}}
The development of the CPPST is geared towards higher education students enrolled in a computing study program. We surveyed both first-year and last-year students and analysed the data as a whole to better represent the entire curriculum. When inspecting the data of first-years and last-years separately, we noticed slight differences in item-total correlation and Cronbach's $\alpha$ values for certain FGS domains. While CUR, MND and KNW items stay the same, many last-year COM items have lower correlations. For example, COM1 and COM3 are about regularly asking feedback from fellow students and peers. Perhaps graduates feel too confident and think they don't need feedback. Perhaps feedback is not perceived as something that can enhance your own creative ideas, but as something obligatory related to grading. To better interpret the results of our pilot study, we aim to perform a focus group not unlike the FGS, but with first-year and last-year students instead of industry experts. Still too little is known about the perception of creativity in the minds of SE students. We argue that, next to these quantitative results, qualitative data can place these differences into context. 

\hypertarget{summary-1}{%
\subsection{Summary}\label{summary-1}}

Our factor analysis yields three major components for CPS in
SE: \emph{Ability}, \emph{Mindset}, and \emph{Interaction}.
These factors somewhat overlap with the aforementioned
engineering scale of \cite{amato2011assessing}. They discovered four factors: cognitive approaches (not unlike our factor \emph{Ability}), cognitive challenges and cognitive
preparedness (somewhat overlapping with our factor \emph{Mindset}), and
impulsivity. The major difference is the absence of any interactive
component, although they also do mention the importance of social aspects
throughout  their paper.  In general, many creativity assessment toolkits lack an
interactive component. Another example is the work of
\citet{denson2015developing}, where creative design in engineering is composed of
technical strength, aesthetic appeal, and originality. It is clear to us that our first
component, \emph{Ability}, can easily be found in other metrics. Some metrics
take \emph{Mindset} into consideration, and almost none \emph{Interaction}.

\vspace{-0.2cm}
\hypertarget{limitations}{%
\section{Limitations}\label{limitations}}

\label{sec:limit}

To reduce treats to validity as much a possible, we closely followed the
recommendations for creative assessment outlined by Barbot et al.
\citep{barbot2019creativity}. For example, they indicate that measuring
creative achievement is perhaps the most objective way to assess
creativity. We therefore formulated the questions in the context of a concrete 
achievement (a programming project), and administered the survey as close as possible to the project deadline.

In general, self-rating is bound to introduce some form of bias. People are
usually inaccurate in assessing their own ``generic creativity'', as
stated by Barbot et al.~We therefore chose to avoid the word ``creativity'' altogether in our survey. Instead, we opted for descriptive, concrete
questions, related to programming as much as possible.  This also sidesteps potential issues due to a fixed mindset on creativity as mentioned by Apiola and Sutinen \citep{apiola2020mindset}.

We believe that the results of this pilot study show promise. The Kaiser-Meyer-Olkin statistic indicates a high factorability, and the Comparative Fit Index and Cronbach's $\alpha$ reliability values confirm that the three-factor model is a good fit. The
three major constructs identified by the exploratory factor analysis are
interesting and explainable, albeit in need of more qualitative context to better understand the perception of students' creativity.

\vspace{-0.2cm}
\hypertarget{conclusion}{%
\section{Conclusion}\label{conclusion}}

\label{sec:concl}

In this paper, we explored the possibility for undergraduate software
developers to self-assess their CPS skills through a survey that
encompasses seven dimensions of creativity, based on existing validated
scales and conducted focus groups. The Creative Programming Problem
Solving Test or CPPST is a survey geared specifically towards computing
education, as we found that existing creativity metrics are either too
broad or too narrow, and do not match well with expectations of software
engineering industry experts.

Principal axis factor analysis grouped the focus group dimensions into
three overarching constructs: (1) \emph{Ability}, (2) \emph{Mindset},
and (3) \emph{Interaction}. We believe that the found constructs can be used as important building blocks for amplifying students' creativity in
computing education. Furthermore, this work contributes towards
computing education research by bringing research on creativity in the
field of cognitive psychology closer to the field of computing
education. The test could be used to gauge the creativity levels of a student assignment.  In future work, we plan to conduct a focus group to gather qualitative data and iterate on the CPPST to further
improve the validity of the self-test. We will also
use it as a pre and post education intervention metric in order to
measure the effectiveness of creativity teaching techniques.

\appendix

\hypertarget{full-reliability-item-statistics}{%
\section{Full reliability item
statistics}\label{full-reliability-item-statistics}}

\label{sec:appendix}

\begin{table}[h!]
  \centering
  \footnotesize
  \caption{Reliability item statistics, calculated individually per domain. Items in bold are excluded after inspecting r.cor.\label{tab:itemstats}}
  {\begin{tabular}{l r r r r r r c}
  \toprule
  item & raw.r & std.r & r.cor & r.drop & mean & sd & invert? \\
  \midrule
CUR1 & 0.018 & 0.0232 & \textbf{0.007} & 0.030 & 3.3 & 0.96 & \\ 
CUR2 & 0.629 & 0.6222 & 0.629 & 0.599 & 3.5 & 0.96 & \\ 
CUR3 & 0.603 & 0.6042 & 0.605 & 0.574 & 3.7 & 0.87 & \\ 
CUR4 & 0.677 & 0.6693 & 0.678 & 0.651 & 3.8 & 0.94 & \\ 
CUR5 & 0.495 & 0.5006 & 0.493 & 0.459 & 3.4 & 0.95 & \\ 
CUR6 & 0.554 & 0.5548 & 0.550 & 0.516 & 3.4 & 1.07 & \\ 
CUR7 & 0.636 & 0.6171 & 0.622 & 0.602 & 3.6 & 1.09 & \checkmark \\ 
CUR8 & 0.553 & 0.5388 & 0.535 & 0.515 & 3.2 & 1.07 & \checkmark \\ 
\midrule
MND1 & 0.257 & 0.2685 & \textbf{0.247} & 0.217 & 3.6 & 0.86 & \\ 
MND2 & 0.263 & 0.2475 & \textbf{0.224} & 0.206 & 3.0 & 1.21 & \\ 
MND3 & 0.663 & 0.6576 & 0.664 & 0.635 & 3.7 & 0.95 & \\ 
MND4 & 0.563 & 0.5560 & 0.553 & 0.524 & 3.6 & 1.09 & \\ 
MND5 & 0.626 & 0.6105 & 0.616 & 0.591 & 3.2 & 1.10 & \checkmark \\ 
MND6 & 0.479 & 0.4534 & 0.450 & 0.434 & 2.5 & 1.13 & \\ 
MND7 & 0.419 & 0.4172 & 0.403 & 0.379 & 3.5 & 0.95 & \checkmark \\ 
MND8 & 0.011 & 0.0014 & \textbf{0.040} & 0.043 & 3.0 & 1.09 & \\ 
\midrule
TCH1 & 0.249 & 0.2366 & \textbf{0.281} & 0.285 & 2.9 & 0.79 & \checkmark \\ 
TCH2 & 0.319 & 0.3194 & 0.299 & 0.279 & 3.4 & 0.88 & \\ 
TCH3 & 0.493 & 0.5116 & 0.502 & 0.467 & 3.7 & 0.68 & \\ 
TCH4 & 0.156 & 0.1589 & \textbf{0.128} & 0.109 & 3.7 & 0.96 & \\ 
TCH5 & 0.298 & 0.3119 & 0.290 & 0.258 & 3.8 & 0.87 & \\ 
TCH6 & 0.296 & 0.3136 & 0.290 & 0.259 & 3.6 & 0.80 & \\ 
TCH7 & 0.187 & 0.1952 & \textbf{0.171} & 0.140 & 3.5 & 0.99 & \\ 
TCH8 & 0.413 & 0.3907 & 0.383 & 0.366 & 2.6 & 1.12 & \checkmark \\ 
\midrule
KNW1 & 0.582 & 0.5903 & 0.593 & 0.555 & 4.0 & 0.80 & \checkmark \\ 
KNW2 & 0.478 & 0.4923 & 0.483 & 0.450 & 3.9 & 0.71 & \\ 
KNW3 & 0.577 & 0.5868 & 0.584 & 0.551 & 3.7 & 0.77 & \\ 
KNW4 & 0.613 & 0.6016 & 0.606 & 0.578 & 3.5 & 1.10 & \\ 
KNW5 & 0.242 & 0.2494 & \textbf{0.226} & 0.199 & 3.4 & 0.90 & \\ 
KNW6 & 0.574 & 0.5547 & 0.555 & 0.535 & 3.2 & 1.14 & \checkmark \\ 
KNW7 & 0.373 & 0.3781 & 0.360 & 0.336 & 3.7 & 0.83 & \\ 
KNW8 & 0.461 & 0.4429 & 0.435 & 0.418 & 3.1 & 1.06 & \\ 
\midrule
COM1 & 0.191 & 0.2063 & \textbf{0.180} & 0.147 & 3.7 & 0.91 & \checkmark \\ 
COM2 & 0.324 & 0.3066 & \textbf{0.286} & 0.267 & 3.2 & 1.25 & \\ 
COM3 & 0.242 & 0.2319 & \textbf{0.206} & 0.191 & 2.9 & 1.08 & \\ 
COM4 & 0.464 & 0.4846 & 0.475 & 0.436 & 4.0 & 0.70 & \\ 
COM5 & 0.372 & 0.3724 & 0.359 & 0.335 & 4.2 & 0.87 & \checkmark \\ 
COM6 & 0.173 & 0.2033 & \textbf{0.181} & 0.143 & 4.1 & 0.62 & \\ 
COM7 & 0.501 & 0.4885 & 0.483 & 0.456 & 3.3 & 1.16 & \\ 
COM8 & 0.301 & 0.3333 & 0.316 & 0.275 & 4.0 & 0.56 & \\ 
\midrule
CTR1 & 0.397 & 0.4186 & 0.405 & 0.369 & 3.9 & 0.68 & \\ 
CTR2 & 0.463 & 0.4497 & 0.438 & 0.418 & 3.3 & 1.12 & \checkmark \\ 
CTR3 & 0.491 & 0.4939 & 0.487 & 0.457 & 3.6 & 0.89 & \\ 
CTR4 & 0.251 & 0.2583 & \textbf{0.234} & 0.214 & 3.3 & 0.79 & \\ 
CTR5 & 0.318 & 0.2941 & \textbf{0.277} & 0.256 & 2.7 & 1.35 & \\ 
CTR6 & 0.276 & 0.2680 & \textbf{0.244} & 0.229 & 3.4 & 1.01 & \checkmark \\ 
CTR7 & 0.457 & 0.4605 & 0.450 & 0.418 & 3.5 & 0.96 & \\ 
CTR8 & 0.408 & 0.4215 & 0.406 & 0.375 & 4.0 & 0.77 & \\ 
\midrule
CRI1 & 0.412 & 0.4363 & 0.426 & 0.382 & 3.7 & 0.73 & \\ 
CRI2 & 0.401 & 0.4208 & 0.405 & 0.370 & 3.5 & 0.73 & \\ 
CRI3 & 0.412 & 0.4142 & 0.400 & 0.370 & 3.8 & 0.99 & \\ 
CRI4 & 0.259 & 0.2659 & \textbf{0.244} & 0.215 & 3.7 & 0.94 & \\ 
CRI5 & 0.131 & 0.1321 & \textbf{0.099} & 0.079 & 3.9 & 1.05 & \checkmark \\ 
CRI6 & 0.181 & 0.1834 & \textbf{0.157} & 0.133 & 3.2 & 0.98 & \\ 
CRI7 & 0.322 & 0.3154 & 0.296 & 0.271 & 3.1 & 1.12 & \checkmark \\ 
CRI8 & 0.222 & 0.2310 & \textbf{0.207} & 0.178 & 3.6 & 0.93 & \\
\bottomrule
  \end{tabular}}
\end{table}

\hypertarget{acknowledgments}{%
\section*{Acknowledgments}\label{acknowledgments}}

We would like to thank Dr.~Barbot's creativity research team of the
Psychological Sciences Research Institute from UCLouvan for their
numerous suggestions and critical view to an early draft of our survey.

\balance
\bibliography{pilot.bib}

\end{document}